\documentstyle[aps,prb,floats,epsf,amssymb]{revtex}

\newcommand{\br}{{\bf r}}

\newcommand{\bv}{{\bf v}}

\newcommand{\bR}{{\bf R}}

\newcommand{\kB}{k_{\rm B}}
\newcommand{\half}{\frac{1}{2}}
\newcommand{\f}{\frac}

\newcommand{\Eq}[1]{Eq.~(\ref{#1})}
\newcommand{\Fig}[1]{Fig.~\ref{#1}}
\newcommand{\Ref}[1]{(\ref{#1})}

\newcommand{\ave}[1]{\overline{#1}}
\newcommand{\eopt}{\epsilon_{\mathrm{opt}}}
\newcommand{\eotp}{\epsilon_{\mathrm{opt}}^{\mathrm{typ}}}
\newcommand{\vopt}{v_{\mathrm{opt}}}
\newcommand{\votp}{v_{\mathrm{opt}}^{\mathrm{typ}}}
\newcommand{\thave}[1]{\overline{\left<#1\right>}}

\begin{document}

\title{Vortex wandering in a forest of splayed columnar defects}

\author{Jack Lidmar, David R.\ Nelson, and Denis A.\ Gorokhov}

\address{
Lyman Laboratory of Physics,
Harvard University,
Cambridge, MA 02138,
USA
}

\date{May 11, 2001}

\wideabs{

\maketitle
\draft

\begin{abstract}
  We investigate the scaling properties of single flux lines in a
  random pinning landscape consisting of splayed columnar defects.
  Such correlated defects can be injected into Type II superconductors
  by inducing nuclear fission or via direct heavy ion irradiation. The
  result is often very efficient pinning of the vortices which gives,
  e.g., a strongly enhanced critical current.  The wandering exponent
  $\zeta$ and the free energy exponent $\omega$ of a single flux line
  in such a disordered environment are obtained analytically from
  scaling arguments combined with extreme-value statistics.  In
  contrast to the case of point disorder, where these exponents are
  universal, we find a dependence of the exponents on details in the
  probability distribution of the low lying energies of the columnar
  defects.  The analytical results show excellent agreement with
  numerical transfer matrix calculations in two and three dimensions.
\end{abstract}


}

\section{Introduction}

Pinning of vortex lines to defects in superconductors plays an
extremely important role in determining their
properties.\cite{Blatter-94}  Motion of flux lines in response to an
applied current induces a voltage and hence dissipation.  Only if the
flux lines remain pinned will the linear resistance be truly zero and
the material exhibit superconductivity.  By introducing controlled
disorder it is possible to significantly improve the critical
currents, temperatures, and fields.  A particularly efficient strategy
is to bombard the sample with heavy ions to produce linear damage
tracks which form optimal pinning centers for the vortex
lines.\cite{civale-91}  When these columnar defects are parallel the
vortices will, below a critical temperature, become localized in a
Bose glass phase that replaces the Abrikosov vortex lattice of the
clean system.\cite{Nelson-Vinokur}
Further improvements were suggested to occur for splayed columnar
defects,\cite{Hwa-93} where the angle mismatch between neighboring
columns should lead to increased barriers for variable range hopping
and therefore an even stronger enhancement of the critical current in
the splay glass phase.  Even more important is perhaps the suppression
of vortex motion due to forced entanglement that is created by the
splay.

There are several ways in which splayed configurations of columnar
defects can be produced.  
A narrow approximately Gaussian angle distribution can be produced by
placing a thin metal foil in front of the sample during irradiation to
defocus the beam of heavy ions as they enter the sample.
Another interesting approach is to induce fission in some of the
nuclei in the material.  The fission products will then go apart in
opposite directions creating damage tracks with a more isotropic
distribution of angles.
Lastly, heavy ion irradiation can be applied at different discrete
angles to create several families of parallel columnar defects.  In
this way it has been possible to increase critical currents by more
than an order of magnitude in many cases.\cite{splay-expr}

The theoretical understanding of the Bose glass, occurring for
parallel columnar defects, rests to a large extent on the mapping of
the vortex line problem to a quantum mechanical system of
(2+1)-dimensional bosons in a random potential, with imaginary time
playing the role of the $z$-axis parallel to the external magnetic
field.  With randomly splayed defects, however, the disorder develops
long range correlations in space and ``time'', and this analogy is
less useful.  In fact, splay disorder leads to logarithmically
divergent phase fluctuations in the boson order
parameter!\cite{Tauber-Nelson}  The properties of {\em individual}
flux lines in such disordered environments are just beginning to be
investigated.\cite{Lehrer-Nelson}


In this paper we focus on the scaling properties of a single flux line
in a sample with many randomly tilted columnar defects.  We consider
the flux line as it enters the superconducting sample on one side at
some arbitrary but fixed position (taken to be the origin), and leaves
on the opposite.  Physically, this fixing of the starting position
could be due to, e.g., surface pinning as in the experimental setup of
\Fig{fig:experiment} below.  The wandering and energetics of the flux
line can be characterized by two exponents defined by the following
relations:
\begin{mathletters}				\label{eq:def-of-exps}
\begin{eqnarray}
&&\thave{ \br^2(z) } \sim z^{2\zeta}, \\
&&\ave{\Delta F^2(z)} \sim z^{2\omega},
\end{eqnarray}
\end{mathletters}
where $\br$ is the transverse position, $z$ is the distance traversed
parallel to the magnetic field, $F$ is the free energy, $\Delta F=
F-\ave{F}$, the overbar denotes a disorder average, and
$\left<\cdots\right>$ a thermal average.

These problems, usually referred to as directed polymers in random
media (DPRM), have received much attention in recent
years,\cite{Halpin-Healy-Zhang-95} with interesting connections to
many other very different physical systems.  The problem can, e.g., be
mapped to the noisy Burgers equation describing a randomly stirred
turbulent fluid, or equivalently, the Kardar-Parisi-Zhang (KPZ) model
of surface growth.
Without disorder, a flux line subject only to thermal fluctuations
will make a diffusion-like random walk as it wanders from the bottom
to the top of the sample with $\zeta=\half$ and $\omega=0$.
Point disorder tends to increase the transverse wandering of the flux
line leading to a non-trivial universal exponent $\zeta >\half$.
(In (1+1) dimensions the exponents are known to be exactly $\zeta=2/3$
and $\omega=1/3$, in (2+1) we have $\zeta\approx 5/8$ and
$\omega\approx 1/4$.\cite{Halpin-Healy-Zhang-95})  In this case
statistical tilt symmetry ensures that $\omega=2\zeta-1$ for arbitrary
$d$.  The fact that $\omega>0$ suggests that energy barriers become
arbitrarily large for large $z$, so that the system is governed by a
zero temperature fixed point, where thermal fluctuations are
irrelevant on long length scales.
Parallel columnar disorder, on the other hand, tends to localize the
flux line, i.e., to reduce its transverse fluctuations.  However, in
search of ever lower pinning energies it will still wander with
non-trivial exponents.\cite{Krug-Halpin-Healy-93}  At zero
temperature the exponents depend on the low energy details in the
disorder distribution, e.g., $\zeta=\omega=\nu/(d+\nu)$ for bounded
distributions of the form \Eq{eq:P(u)} below, while at finite
temperature the wandering is slightly sub-ballistic, with
$\ave{|\br(z)|} \sim z/(\ln z)^\gamma$, $\gamma = 1+2/d$, where $d$ is
the number of dimensions perpendicular to
$z$.\cite{Krug-Halpin-Healy-93}
For {\em splayed} columnar defects one might expect an enhancement of
this wandering since the flux lines try to follow the randomly tilted
defects.

The path a flux lines takes inside a superconductor is not easy to
measure experimentally.  What can be measured is the positions at
which the flux line enter and leave the superconductor, e.g., using
double-sided decoration experiments.  In \Fig{fig:experiment} we
propose an experimental setup which would allow the endpoints $\br(z)$
of the flux line to be measured for several different disorder
realizations of varying height $z$, using a single piece of
superconductor.  Short parallel columnar defects on one side of the
sample provide a set of known starting points for the vortices as they
enter the forest of splayed columnar defects and traverse the sample.
At low magnetic field it should be possible to approximately match
entry and exit points via double-sided flux decorations.\cite{Lieber}
The variable length parallel defects simulate the effect of varying
sample thickness.

\begin{figure}
\centerline{ \epsfxsize=1\linewidth \epsfbox{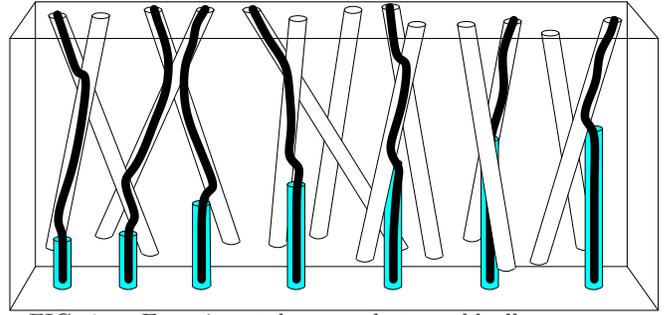} }
\caption{
Experimental setup that would allow measurement of the wandering
exponent $\zeta$.  Parallel columnar defects of varying depth are
introduced from the bottom of the sample in order to provide a strong
pinning mechanism to fix the entry positions of the flux lines.
Splayed columnar defects are introduced from the other side, and the
relative transverse distance between the starting point and the exit
point of the flux line could then be measured, e.g., in double sided
decoration experiments, or using magneto-optical imaging.
A similar setup could also be used to investigate the vortex
wandering in point disordered samples.
}
\label{fig:experiment}
\end{figure}

Apart from the intrinsic interest in the behavior of a single flux
line in a disordered superconductor, the physics of a single vortex in
the presence of splay is relevant for understanding properties at low
magnetic fields, where the flux density is low.
It affects, e.g., the constitutive relation $B(H)$ just above the
$H_{c1}$-line, where a balance between magnetic field, pinning energy,
and repulsive interactions gives $B \sim (H-H_{c1})^\beta$, with
$\beta = d\zeta/(1-\omega)$.\cite{Nattermann-Lipowsky,Lehrer-Nelson}
This relation was recently measured in experiments in a (1+1)
dimensional geometry for samples with point disorder,\cite{Bolle-99}
confirming the prediction $\beta=1$.
Furthermore, the non-linear current-voltage (IV) characteristics at
low temperatures and currents is determined by the depinning of single
flux lines from the defects.  If we assume that the free energy {\em
barriers} scale in the same way as the sample-to-sample free energy
{\em fluctuations}, an excitation of length $z$ and width $\ell \sim
z^\zeta$ would cost pinning energy $\Delta E \sim z^\omega$ but gain
$J \ell z$ due to the Lorenz force from the current density $J$, so
that the free energy barrier to overcome is $\Delta F(J,z) = k_1
z^\omega - k_2 J z^{1+\zeta}$.  Optimizing with respect to $z$ gives
$\Delta F(J) = A J^{-\mu}$, where $\mu=\omega/(\zeta+1-\omega)$.
Thermally activated creep over the free energy barriers then gives
rise to an electric field ${\mathcal E} \sim
\exp\left[{-(J_0(T)/J)^\mu}\right]$, with $J_0(T)=(A/T)^{1/\mu}$.
Finally, the study of single flux lines also allows estimates of
entanglement lengths of interacting vortices, which may shed light on
the still mysterious properties of the splay glass phase.

While the single vortex pinning from point disorder and parallel
columnar defects has received much attention, the case with splayed
columnar disorder has only been addressed quite
recently,\cite{Lehrer-Nelson} and only for the special case of a
``nearly'' isotropic distribution of splay angles.  There it was
argued that, for very wide angle distributions, the problem would
reduce to a flux line in a disorder landscape with long range spatial
transverse correlations but no correlations along the magnetic field.
Explicitly, the Fourier transform of the disorder correlator for
general splayed configurations is $\Delta(k_\perp,k_z)=\Delta_0
f(k_z/v_0 k_\perp) / v_0 k_\perp$, for small $k_\perp, k_z$, where
$v_0$ is the characteristic width of the distribution of slopes.
``Nearly isotropic'' splay refers to the limit $v_0 \to \infty$
(keeping $\Delta_0/v_0$ fixed), giving $\Delta(k_\perp,k_z) \propto
1/k_\perp$, while a truly isotropic distribution would give
$\Delta(k_\perp,k_z) \propto 1/\sqrt{k_\perp^2 + k_z^2}$.  Neglect of
the $k_z$-dependence is at least plausible in the case $\zeta < 1$,
since the width grows like $v_0 \sim [{\mathrm{length}}]^{1-\zeta}$
under renormalization and therefore would diverge to infinity on large
length scales.  Subject to these assumptions, the flux line problem
can then be mapped to a noisy Burgers equation with spatially
correlated noise,\cite{Lehrer-Nelson} for which the {\em universal}
results\cite{Frey} $\zeta=3/(3+d)$ and $\omega=(3-d)/(3+d)$ have been
conjectured.  In this approach, however, higher order correlation
functions of the disorder are neglected.

Here we consider splayed columnar defects for general angle
distributions using a real space approach, similar in spirit to
Ref.~\onlinecite{Krug-Halpin-Healy-93} for parallel columnar defects.  This
leads to a rather different physical picture, where the important
feature is that the flux line keeps following columnar pins with lower
and lower energy (and smaller and smaller slopes).  We find new
exponents that depend on the tail of the distribution of the energies
of the columns.  The sensitivity to low energies suggests that higher
correlation functions of the disorder, which were missing in the
previous treatment, may be important.  The predictions for $\zeta$ and
$\omega$ of the different approaches are compared in
Table~\ref{tab:exponents}.
We have been able to confirm the results of the real space approach
using numerical transfer matrix calculations in (1+1) and (2+1)
dimensions.

In Sec.~\ref{sec:analytic} we use a combination of scaling arguments
and extremal statistics to derive analytic expressions for the
wandering and energy exponents.  We also discuss crossover effects and
other complications, such as inhomogeneities of the columnar defects
(fragmentation), and the influence of temperature.
Section~\ref{sec:transfer} describes our numerical work.  We introduce
a model well suited for transfer matrix calculations of
superconductors containing splayed columnar defects, discuss finite
size effects, and present our results.
In Sec.~\ref{sec:future} we review some open questions.

\section{Analytic arguments}

\label{sec:analytic}

\subsection{The model}

The Hamiltonian of the flux line consists of an elastic energy and
an interaction with the disorder,
\begin{equation}				\label{eq:continuum-H}
  H = \int_0^{L_z} \!\!\!\! dz \left\{ 
	\f{\gamma}{2} \left(\f{d\br}{dz}\right)^2 + U(\br(z),z)\right\},
\end{equation}
where $\br$ is the transverse $d$-dimensional position, $z$ the
distance parallel to the magnetic field, and $\gamma$ is the line
tension.  Splayed columnar defects are distributed throughout the
system, giving a disorder potential $U(\br,z) = \sum_i U_i(\br-\bv_i z
-\br_i^0)$, with random slopes $\bv_i$ and random potential wells
$U_i$.  The $\br_i^0$'s are the positions where the columns cross the
plane $z=0$.  We are interested in the low-temperature regime where
the flux line is tightly bound to the defects.  The binding energy per
length of a flux line pinned to a column is
then\cite{Nelson-Vinokur,Hwa-93} $\epsilon_i = \f{\gamma}{2} \bv_i^2 +
u_i$, where $u_i \approx U_i({\bf 0}) + c T^2/(2\gamma b_i^2)$, $b_i$
being the radius of the defects, and $c$ a constant of order unity.

The probability distribution $P(\epsilon)$ of the binding energies
depends on details of the injection process in the material.  Below we
will need information about the low energy tails of this distribution.
Given the probability distributions $P_u(u)$ of the depths and
$P_v(\bv)$ of the slopes, which we assume to be independent, we get
for the probability of having total energy per length $\epsilon$ the
expression
\begin{eqnarray}
 P(\epsilon)&=& \int du \int d^d\bv P_u(u) P_v(\bv)
    \delta(\epsilon - u - \f{\gamma}{2} \bv^2) \nonumber \\
  &=& \int d^d\bv P_u(\epsilon - \f{\gamma}{2} \bv^2) P_v(\bv).
\end{eqnarray}
In any realistic case the potential distribution will be bounded, and
for convenience we will use
\begin{equation}					\label{eq:P(u)}
  P_u(u) = \left\{
\begin{array}{ll}
   \f{\nu u^{\nu-1}}{\Delta^\nu}, & \textrm{ if } \; 0 \le u \le \Delta \\
   0, & \textrm{ otherwise}.
\end{array} \right.
\end{equation}
A uniform bounded distribution is perhaps the most natural choice and
corresponds to $\nu=1$, while the limit $\nu\to 0$ gives a delta
function, i.e., all columns have the same depth.  In the general case
$\nu > 0$.
We have further chosen the energy scale such that zero corresponds to
the deepest possible pin.
We assume a continuous symmetric distribution, $P_v(\bv)$, of the slope of the
columnar defects, with a finite value for $\bv \to 0$ and a
characteristic width $v_0$.
Low energies turn out to be the most important ones, and we find
\begin{equation}				\label{eq:energyscaling}
  P(\epsilon)d\epsilon \propto \left( \epsilon \over \Delta \right)^\nu
	\left( \epsilon \over \f{\gamma}{2}v_0^2 \right)^{d/2}
	{d\epsilon \over \epsilon},
 \qquad 0 < \epsilon \lesssim \f{\gamma}{2}v_0^2\,, \;\; \Delta ,
\end{equation}
i.e., a power law behavior $P(\epsilon) \sim \epsilon^{\alpha-1}$ with
\begin{equation}					\label{eq:alpha}
 \alpha=\f{d}{2} + \nu .
\end{equation}
Under the assumptions made above this relation should hold in the tail
even in the presence of weak correlations between $P_u(u)$ and
$P_v(v)$.

A complication may arise from the fact that the defects are not
necessarily uniform along their extension.  Since the heavy ions
creating the damage tracks lose energy to the material as they travel
through it, there may be a systematic dependence of the diameter and
angle on $z$.
The tracks formed by low-energy ions are furthermore often fragmented.
We will neglect these effects for now and return to them in the
discussion below.  With a good model of these processes it should be
possible to extend the arguments given below to take into account
these facts.
In practice $P(\epsilon)$ could have several regimes with an
approximate power law behavior, leading to crossovers between different
scaling regimes before the asymptotic result determined by
\Eq{eq:alpha}, is reached (see below).

\subsection{Scaling theory from extremal statistics}

\label{sec:extremal}

As a flux line travels through the sample at low temperature it seeks
out pins with lower and lower energy.  At a given distance $z$ the
flux line has had a chance to find the optimal columnar pin within a
region of size $\ell^d \times z$, with $\ell \sim z^{\zeta}$.  In this
region there are on average $N=\rho\ell^d$ pins, where $\rho$ is the
areal density perpendicular to $z$.  For a given length scale $\ell$
the physics will be dominated by the pin with lowest energy among
these $N$ pins.  The typical energy $\eotp$ per unit length of this
optimal column can then be estimated using arguments of extremal
statistics \cite{extremal-statistics} if $N$ is large enough, i.e.,
for large $z$.  Upon introducing the probability that a pin has an
energy {\em less} than $\epsilon$,
\begin{equation}
F(\epsilon)=\int_{-\infty}^\epsilon
P(\epsilon')d\epsilon' \propto \f{\epsilon^\alpha}{
\Delta^\nu \left( \gamma v_0^2 / 2 \right)^{d/2} 
},
\end{equation}
we have
\begin{equation}					\label{eq:extr-der}
  {\mathrm{Prob}}(\eopt > \epsilon ) = (1 - F(\epsilon))^N
	\to e^{-NF(\epsilon)}, \; \textrm{  for } N \textrm{ large}.
\end{equation}
For a typical energy of the optimal column we have $NF(\eotp)\sim1$,
i.e.,
\begin{equation}
  \eotp \sim N^{- \f{1}{\alpha}} \sim z^{-\f{\zeta d}{\alpha} }.
\end{equation}
The total energy of the whole optimal path will then scale as
\begin{equation}
  E_{\mathrm{typ}} \sim \int_0^z \eotp(z')dz' 
  \sim z^{1-\f{\zeta d}{\alpha}} \sim z^{\omega}.
\end{equation}
Thus we have obtained the relation
\begin{equation}				\label{eq:omega-zeta-relation}
\omega = 1 - \zeta d / \alpha
\end{equation}
between the energy exponent and the wandering exponent.  Note the
dependence (through $\alpha$ and \Eq{eq:alpha}) on the detailed
behavior of $P_u(u)$ at low energies.

Now we will turn to the determination of $\zeta$ itself.  Assuming
that the flux line is mostly localized along the columns with lowest
energy, we only need to know how the slope $\vopt$ of the optimal pin
encountered decreases with $z$.  The probability of $\vopt$ is, for
large $N$, given by
\begin{eqnarray}
\lefteqn{ P_{\vopt}(\vopt)d^d\bv_{\mathrm{opt}} = }&& \nonumber \\
 &\qquad&  N \int d\epsilon P_u(\epsilon-\f{\gamma}{2}\vopt^2) P_v(\vopt) e^{-N
  F(\epsilon)} d^d\bv_{\mathrm{opt}}.
\end{eqnarray}
A simple calculation shows that $\votp \sim \sqrt{2 \eotp/\gamma} \sim
N^{-1/2\alpha} \sim z^{-\zeta d / 2 \alpha}$ for $\eotp \lesssim \gamma
v_0^2/2$, and for the typical value of the endpoint $\br(z)$ of the
flux line at distance $z$ we then obtain
\begin{equation}
  r_{\mathrm typ}(z) \sim \int_0^z \votp(z')dz' 
	\sim z^{1-\zeta d /2 \alpha} \sim z^\zeta .
\end{equation}
Thus $\zeta=1/(1+d/2\alpha) = 2\alpha/(2\alpha+d)$, and via
\Eq{eq:omega-zeta-relation}, $\omega = (2\alpha-d)/(2\alpha+d)$.  From
\Eq{eq:alpha} we then obtain our final result,
\begin{mathletters}				\label{eq:exponents}
\begin{eqnarray}
  \zeta  &=& \f{d/2+\nu}{d+\nu} \\
  \omega &=& \f{\nu}{d+\nu}.
\end{eqnarray}
\end{mathletters}
Remarkably, these exponents satisfy for all $d$ and $\nu$ the relation
$\omega=2\zeta-1$ valid for systems obeying statistical tilt symmetry,
\cite{Halpin-Healy-Zhang-95} even though strictly speaking, there is
{\em no} exact tilt symmetry in the present case.

Since $0<\nu<\infty$, we always have $\half \le \zeta \le 1$ and $0
\le \omega \le 1$.  For a uniform bounded disorder distribution
($\nu=1$) we get $\zeta=3/4$, $\omega=1/2$ in (1+1)D and $\zeta=2/3$,
$\omega=1/3$ in (2+1)D, leading to an enhanced wandering compared to
point disorder.  If all splayed column well depths are identical, $\nu
\to 0$ giving $\zeta=\half$ and $\omega=0$.  This means that the $T=0$
fixed point becomes marginal and thermal fluctuations (as well as
point disorder) could become important.  For $d \to \infty$ the
exponents smoothly attain their thermal values $\zeta=\half$, $\omega
= 0$.  If $P_u(u)$ is replaced by a Gaussian distribution (thus
allowing arbitrarily low pinning energies) we find $\zeta = \omega =
1$, described by the $\nu\to \infty$ limit, but in this case there may
also be logarithmic corrections.  A distribution with power-law tails,
$P_u(u) \sim |u|^{-\nu-1}$, $u<-1$, would give $\zeta=1$ and
$\omega=1+d/\nu$.
In Table~\ref{tab:exponents} we summarize the exponents found for
splayed columnar defects and compare them with the exponents for other
kinds of disorder.  Comparing with the results found in
Ref.~\onlinecite{Lehrer-Nelson} for a disorder distribution with long range
transverse correlations, $\Delta(k)\sim 1/k_\perp$, the exponents
agree for the special case $d=1$, $\nu=1$, but disagree in general.

Another interesting quantity to look for is how often the flux line
changes direction.  Let us define the total number of intercolumn
jumps the flux line has made up till $z$ as $n_z$.  The turning
probability in an infinitesimal interval $[z,z+dz]$ is given by the
number of new columns explored times the probability of any of them
being lower in energy than the current energy, $dn_z \sim F(\eotp) dN
\sim dN/N$, which gives
\begin{equation}
  n_z \sim \ln N(z) \sim \ln z.
\end{equation}
From this follows that the spacing along $z$ between two turns
increases like $\Delta z \sim z$ on average.

\begin{table}
\caption{Summary of wandering and energy exponents for some selected
values of $\nu$ and for other kinds of disorder.}
\label{tab:exponents}
\begin{tabular}{lcccc}
& \multicolumn{2}{c}{$d=1$} & \multicolumn{2}{c}{$d=2$} \\
\tableline
& $\zeta$ & $\omega$ & $\zeta$ & $\omega$ \\ 
\tableline
Splayed columnar defects ($\nu=0$)
& $1/2$ & $0$ & $1/2$ & $0$ \\
Splayed columnar defects ($\nu=1$)
& $3/4$ & $1/2$ & $2/3$ & $1/3$ \\
Splayed columnar defects ($\nu=2$)
& $5/6$ & $2/3$ & $3/4$ & $1/2$ \\
$1/k_\perp$-correlator \cite{Lehrer-Nelson}
& $3/4$ & $1/2$ & $3/5$ & $1/5$ \\
Parallel columns ($\nu=1$) \cite{Krug-Halpin-Healy-93}
& $1/2$ & $1/2$ & $1/3$ & $1/3$ \\
Point disorder
& $2/3$ & $1/3$ & $\sim 5/8$~~~ & $\sim 1/4$~~~ \\
Splay with fragmentation
& $3/4$ & $1/2$ & $\sim 5/8$?~~ & $\sim 1/4$?~~ \\
\end{tabular}
\end{table}

\subsection{Crossover effects and scaling regimes}

\label{sec:Crossover}

The results above hold asymptotically for large $z$ and depend only on
the parameter $\nu$ in the pinning energy distribution.  Realistic
pinning energy distributions can be expected to be more complicated
than the simple form \Eq{eq:P(u)} used above and this may lead to
various crossover effects between different scaling regimes, depending
on the energy scale probed at a given length scale.  This may
effectively lead to a $z$-dependent parameter $\nu$, and may also
disrupt the relation $\votp \sim \sqrt{\eotp}$, and hence the relation
$\omega=2\zeta-1$ over certain length scales, although asymptotically
the behavior should be that of the previous section.

For the bounded distributions used above, \Eq{eq:P(u)}, the asymptotic
behavior sets in when $\eotp \lesssim \min \left\{\Delta, {\gamma
v_0^2 /2}\right\}$, or, using $N(z) \sim 1/F(\eotp)$,
\begin{equation}				\label{eq:asymp-cross}
 N(z) \gg
  \max\left\{ 
  \left( \Delta \over \f{\gamma}{2} v_0^2 \right)^\nu,
  \left( \f{\gamma}{2} v_0^2 \over \Delta \right)^{d/2}
 \right\},
\end{equation}
where $N(z)$ is the number of columns explored up till $z$.
Finally, we expect behavior typical of parallel columnar defects when
$z \ll a_\perp / v_0$, where $a_\perp = \rho^{-1/d}$.  As $v_0 \to 0$
this crossover scale diverges as expected.

\subsection{Fragmented columnar defects}

\label{sec:Fragmentation}

\begin{figure}
\centerline{ \epsfxsize=0.8\linewidth \epsfbox{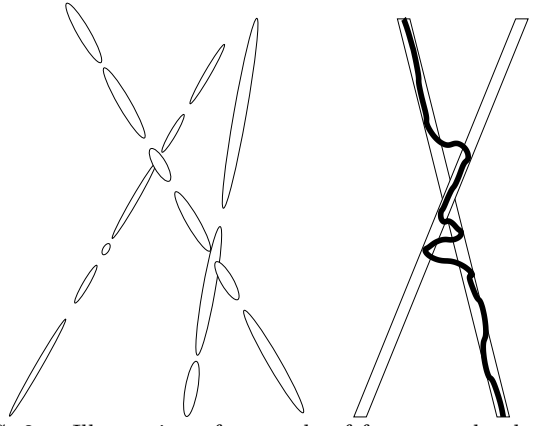} }
\caption{
Illustration of a couple of fragmented columnar defects (left) and a
flux line which fluctuates back and forth between two crossing
columnar defects at finite temperature (right).  The entropic
contribution to the free energy from the crossing points, which appear
randomly along the defects, act very much like fragmentation.
}
\label{fig:frag}
\end{figure}

In some cases the columnar defects created by the ion irradiation (or
fission fragments) are highly nonuniform along their axis (See
\Fig{fig:frag}). \cite{Ertas}  This fragmentation depends on, e.g.,
the kinetic energy and the type of ions used.  In this subsection we
discuss the influence of this fragmentation on the flux line
wandering.

We first consider the stability of the results derived in
Sec.~\ref{sec:extremal}.  With fragmentation the energy of a flux line
pinned to a columnar defect will be subject to two kinds of disorder.
As before there is a random $z$-independent part constant along the
columns, given by \Eq{eq:P(u)}.  We assume that fragmentation
superimposes a random short range correlated energy modulation along
the columns with zero mean.

Let us now consider a segment of a flux line pinned to a defect over a
length $\Delta z \gg a$, where $a$ is the microscopic size of the
fragments.  The fluctuation in the average energy per unit length then
gets a contribution $\sim \epsilon_f=f \sqrt{a/\Delta z}$ from the
fragmentation, where $f$ is the root mean square strength of the
disorder.  Fragmentation should be relevant only if this energy
exceeds the fluctuations in the constant part of the energy.  The
density of columns with energy less than this energy is $\rho_< = \rho
F(\epsilon_f)$, and their mean transverse distance $\ell \approx
(\rho_<)^{-1/d}$.  The mean longitudinal distance between crossings
(or close encounters) of these low energy columns is then $\Delta z^*
\approx \ell / v^*$, where the $v^*$ is the maximum typical slope,
$v^* \sim \sqrt{ 2 \epsilon_f /\gamma } \sim \sqrt{2f/\gamma} \left(a
/\Delta z \right)^{1/4}$.  Introducing the mean transverse spacing of
defects $a_\perp = \rho^{-1/d}$ we find
\begin{equation}
  \Delta z^* \sim \f{a_\perp}{v_0} 
  \f{\f{\gamma v_0^2}{2} \, \Delta^{\nu/d}}{f^{1+\nu/d}}
  \left( \Delta z \over a \right)^{1+\nu/d \over 2}.
\end{equation}
Fragmentation can only be important if $\Delta z^* < \Delta z$, i.e.,
on large length scales if $\nu < d$, and otherwise on short length
scales, with the crossover scale given by the requirement $\Delta z^*
\approx \Delta z$,
\begin{equation}				\label{eq:cross-over}
  \Delta z_{\mathrm{crossover}} \approx a \left( \f{a_\perp}{v_0 a} \f{
  \f{\gamma v_0^2}{2} \Delta^{\nu/d} }
  {f^{1+\nu/d}} \right)^\f{2}{1-\nu/d}.
\end{equation}

What happens when the fragmentation dominates, i.e., on long length
scales for $\nu < d$ or on short length scales for $\nu > d$?
Although the disorder along the columns is correlated only over the
microscopic scale $a$ of the fragments the problem still has long
ranged correlations in that the disorder is restricted to the tracks
of the columns.
There are (at least) two possible scenarios.  If the point-like
contribution to the disorder from many columnar fragments is strong
enough to completely delocalize the flux line, this would presumably
lead to a crossover to the physics dominated by point disorder, with a
corresponding change in the critical exponents.  However, if the
fragmentation is weak, the flux line could remain pinned to the best
fragmented column available in the region explored, occasionally
making excursions to take advantage of favorable fluctuations in the
disorder.
In Ref.~\onlinecite{Balents-Kardar} the depinning of a flux line from a
single columnar defect in the presence of point disorder was studied.
These authors found a transition from a localized to a delocalized
state for $d>1$ as a function of disorder strength, while the case
$d=1$ was marginal with the flux line pinned by an arbitrarily weak
attractive column.
In analogy with these results we do not expect the first scenario to
be realized for $d=1$, whereas in higher dimensions either scenario is
{\em a priori} possible and a phase transition between the two cannot
be excluded.

We now take a closer look at the second scenario of {\em weak}
fragmentation.  The flux line will then effectively see splayed
columnar defects with a Gaussian distribution of pinning energies with
variance $\sim f^2{a/\Delta z}$, that explicitly depends on the
distance $z$ (recall that $\Delta z \sim z$).  Repeating the extremal
statistics calculation of Sec.~\ref{sec:extremal} for this choice of
pinning energies leads to $\eotp \sim f\sqrt{a/\Delta z}$ up to
logarithmic corrections.  The typical optimal slope is $\votp \sim
\sqrt{2\eotp/\gamma}$ again ignoring logarithms.  The total energy of
the flux line then scales as $ E_{\mathrm{typ}}(z) \sim z^{1/2}$, and
the endpoint position as $r_{\mathrm typ}(z) \sim z^{3/4}$, so that
\begin{mathletters}					\label{eq:frag}
\begin{eqnarray}
 \zeta  &=& 3/4 \\
 \omega &=& 1/2,
\end{eqnarray}
\end{mathletters}
when weak fragmentation dominates.

\subsection{Finite temperature}
\label{sec:Finite-T}

We have assumed thus far that temperature was low enough to be ignored
and the problem could be analyzed solely in terms of energetics.  As
the temperature is increased entropy becomes more important.  At
relatively low temperatures the main contribution will come from
regions where two or more columns come closer than $\ell_T \approx
T/\sqrt{2\gamma \Delta U}$, where $\Delta U$ is the energy difference
between a columnar defect and the surrounding material.  The flux line
will then gain entropy by fluctuating back and forth between these
columns \cite{Hwa-93} (See \Fig{fig:frag}).  Since these crossing
points occur randomly along the splayed columns the situation becomes
very similar to that of fragmented columnar defects at zero
temperature.  Therefore, we expect the argument of the previous
subsection to apply, i.e., temperature will be important whenever
fragmentation would be important, and the wandering and energy
exponents should be the same as for that case.  If the temperature is
increased even further the flux line will no longer be pinned to
individual defects, but rather collections of them.  In this case the
situation becomes far more complicated, and beyond the scope of the
present paper.

\section{Transfer matrix calculations}

\label{sec:transfer}

To check the analytic arguments in Sec.~\ref{sec:analytic}, we have
performed numerical transfer matrix calculations.  Usually transfer
matrix calculations of directed polymers are performed on a model
defined on a regular lattice, but for the present problem it is
difficult to account for the varying splay angles using this approach.
Instead we have developed a model on an irregular random lattice
defined by a particular configuration of randomly splayed columnar
defects.

\subsection{Model}

The disorder landscape consists of $N$ splayed columnar defects
embedded randomly in a system of size $L^d \times L_z$.  The
transverse positions of the columnar defects are given by
\begin{equation}
  \br_i(z) = \br^0_i + \bv_i z,
\end{equation}
where the $\br^0_i$ are distributed uniformly in the plane $z=0$, and
the $\bv_i$'s are chosen randomly with some distribution
$P(\bv)d^d\bv$.  We also assign a random energy cost $u_i \in
[0,\Delta]$ to each column, with probability $P_u(u)du$ from
\Eq{eq:P(u)}.  Periodic boundary conditions are employed so that we
always have $\br_i(z)\in[-L/2,+L/2]^d$.  In addition we use a
discretization in the $z$-direction with lattice constant $a_z=1$.

\begin{figure}
\centerline{ \epsfxsize=0.8\linewidth \epsfbox{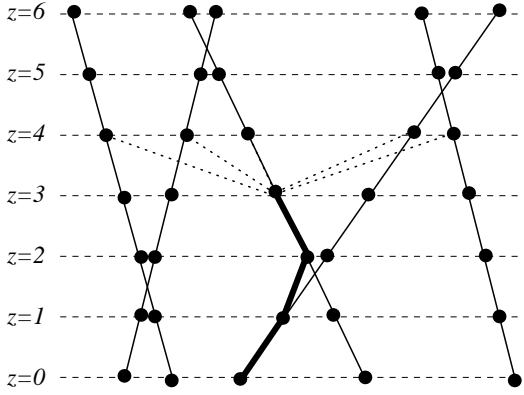} }
\caption{
Discretization used for transfer matrix calculations in (1+1)
dimensions.  At each step the flux line has the possibility to jump to
anyone of the other columnar pins at the next constant-$z$ section.  A
similar picture obtains in (2+1) dimensions.
}
\label{fig:trans}
\end{figure}

A flux line enters the system at $\bR_0=({\mathbf{0}},0)$ and leaves
at some $\bR=(\br,L_z)$.  The model is defined by restricting the
positions of the flux line to the columnar defects.  The space between
pins is excluded, except to allow jumps from one column to another
(see \Fig{fig:trans}).
The flux line can then be parameterized by the sequence $\left\{ i_z
\right\}_{z=0}^{L_z}$ of columns visited as it traverses the sample
from bottom to top.  The actual path a vortex takes is given by
$\br_{i_z}(z)=\br^0_{i_z} + \bv_{i_z}z$, and
the total energy of the flux line is given by
\begin{equation}				\label{Hamiltonian}
  H = \sum_{z=0}^{L_z-1} \Delta E_{i_{z+1},i_z}(z),
\end{equation}
where $\Delta E_{ij}(z)$ is the energy of the (straight) line segment
between $\br_i(z+1)$ and $\br_j(z)$.  We take
\begin{equation}				\label{segment}
  \Delta E_{ij}(z) = \f{\gamma}{2} \left[ \br_i(z+1)-\br_j(z) \right]^2 + u_i,
\end{equation}
where the first term is the elastic energy (the distance appearing in
brackets is understood to be the shortest distance using the periodic
boundary conditions) and the second is a random site energy of column
$i$, consistent with the continuum model \Eq{eq:continuum-H}.  More
elaborate forms could be used but will presumably not change any
universal properties.  Fragmentation of the columnar defects is
modeled below by adding an uncorrelated uniformly distributed term
$f_i(z) \in [0,f]$ to \Eq{segment}.

\subsection{Recursion relations}

The partition function for the flux line now obeys the recursion
relation
\begin{equation}				\label{recursion}
  Z_i(z+1) = \sum_{j=1}^N e^{-\beta \Delta E_{ij}(z)} Z_j(z),
\end{equation}
where $\beta=1/T$ is the inverse temperature, and we use units such
that $\kB=1$.  At zero temperature this reduces to an optimization
problem for the total energy of the path:
\begin{equation}				\label{recursion-0}
  E_i(z+1) = \min_j \left\{ E_j(z) + \Delta E_{ij}(z) \right\}
\end{equation}
The sum in \Eq{recursion} and minimization in \Eq{recursion-0} are
over all possible columns, thus jumps between columns are not
limited to just nearest neighbors, but instead restricted by the
elastic line tension.
These relations are easily iterated on a computer, and averages are
then calculated from
\begin{equation}
  \thave{O(z)} = \ave{\f{ \sum_i O_i(z) Z_i(z)}{ \sum_i Z_i(z) }}.
\end{equation}
The free energy is given by $F(z) = -T \ln Z(z)$, where $Z(z)=\sum_i
Z_i(z)$.  (At zero temperature the free energy is of course equal to
the energy.)  The overbar denotes the quenched average over different
disorder realizations.

\subsection{Finite size scaling}

In a system of finite transverse size $L$, we expect that the mean
square fluctuations of the endpoint will obey
\begin{equation}				\label{eq:collapse-x2}
  \thave{\br^2(z)} = L^2 C(z/L^{1/\zeta}),
\end{equation}
while the free energy fluctuations satisfy
\begin{equation}				\label{eq:collapse-f2}
  \ave{\Delta F^2(z)} = L^{2\chi} f(z/L^{1/\zeta}),
\end{equation}
where $\chi= \omega/\zeta$.  For small arguments the functions $C$ and
$f$ should become power-laws, such that the right hand sides become
independent of $L$.  In the opposite limit of large arguments $C(x)
\to d/12$, and $f(x) \sim x^2$.\cite{foot}
The crossover happens at some value of $x=x_c$, i.e., for $z = x_c
L^{1/\zeta}$.  Thus, when $\thave{\br^2}/L^2$ and $\ave{\Delta
F^2}/L^{2\chi}$ are plotted against $z/L^{1/\zeta}$, the data for
different $L$ should collapse onto a single curve (at least for large
enough $z$).

\subsection{Results and discussion}

\begin{figure}
\centerline{ \epsfxsize=1\linewidth \epsfbox{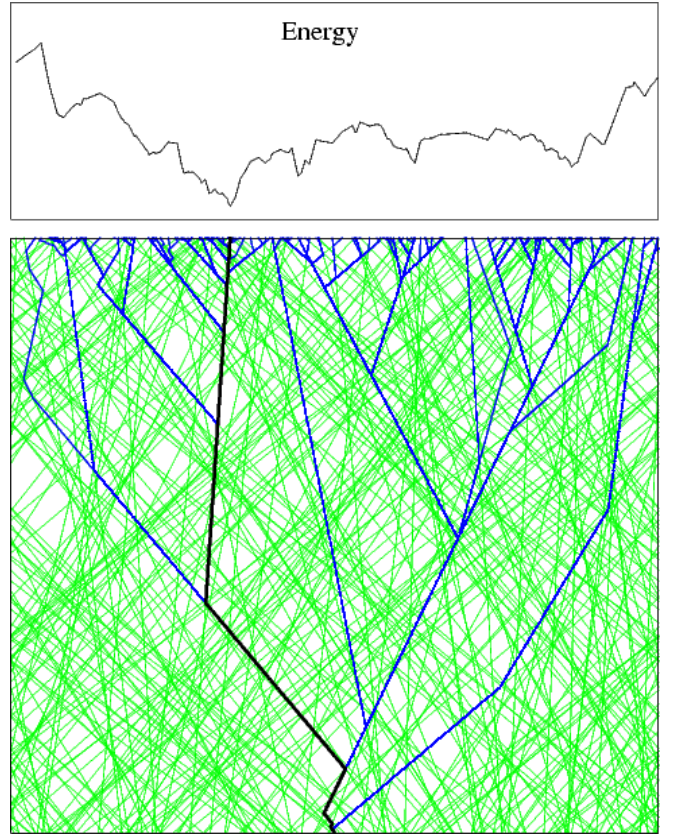} }
\caption{
Example of the optimal paths with variable endpoints in a particular
disorder realization in (1+1) dimensions.  The light gray lines
represent the randomly splayed columns with random energies given by
\Eq{eq:P(u)} with $\nu=1$.  The solid lines are the optimal paths if
the endpoint is restricted to a given position.  The thickest line is
the optimal path among all of them, i.e., without any restrictions.
The upper panel shows the corresponding energies as a function of
endpoint position.
}
\label{fig:path}
\end{figure}

The recursion relations \Eq{recursion} and \Ref{recursion-0} were
solved numerically and averaged over 2000-10000 disorder realizations
to get small error bars.  
Figure~\ref{fig:path} shows an example of the optimal path
(corresponding to $T=0$) of a flux line for a particular disorder
realization.
The number of defects were taken to be $N = L^d$ so that the mean
spacing between defects is unity.  The line tension is set to unity,
equivalent to measuring all energies in units of $\gamma$.  We further
choose the distribution of tilts $P(\bv)$ uniform in $[-v_0,+v_0]^d$.
Apart from the shape of the tail of the pinning potential, given by
the exponent $\nu$, the asymptotic behavior on long length scales
should be independent of $\Delta$ and $v_0$.  However, these values
will influence the length scales where the system crosses over to the
asymptotic regime.  Since the system sizes possible to study
numerically are limited, care must be taken in choosing the values of
these parameters to reduce crossover effects, otherwise inaccurate
results can easily be obtained.  The analysis of
Sec.~\ref{sec:Crossover} suggests that the crossover length scales are
minimized when $\gamma v_0^2/2 \approx \Delta$.  With this in mind we
usually put $v_0=1/2$, $\Delta=0.125$ and sometimes $v_0=1/4$,
$\Delta=0.03$.

\begin{figure*}
\centerline{
\epsfxsize= 0.5 \linewidth \epsfbox{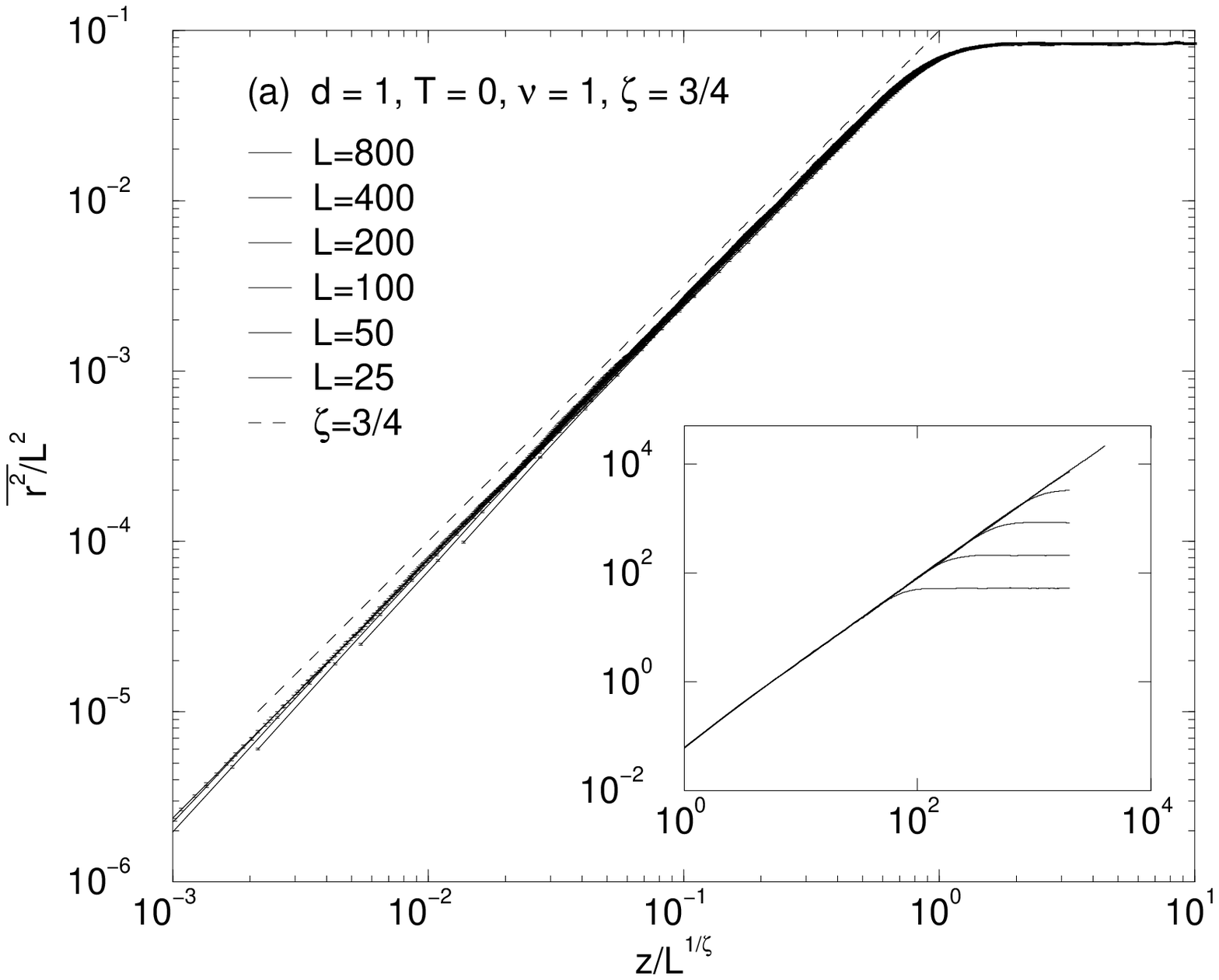}
\epsfxsize= 0.5 \linewidth \epsfbox{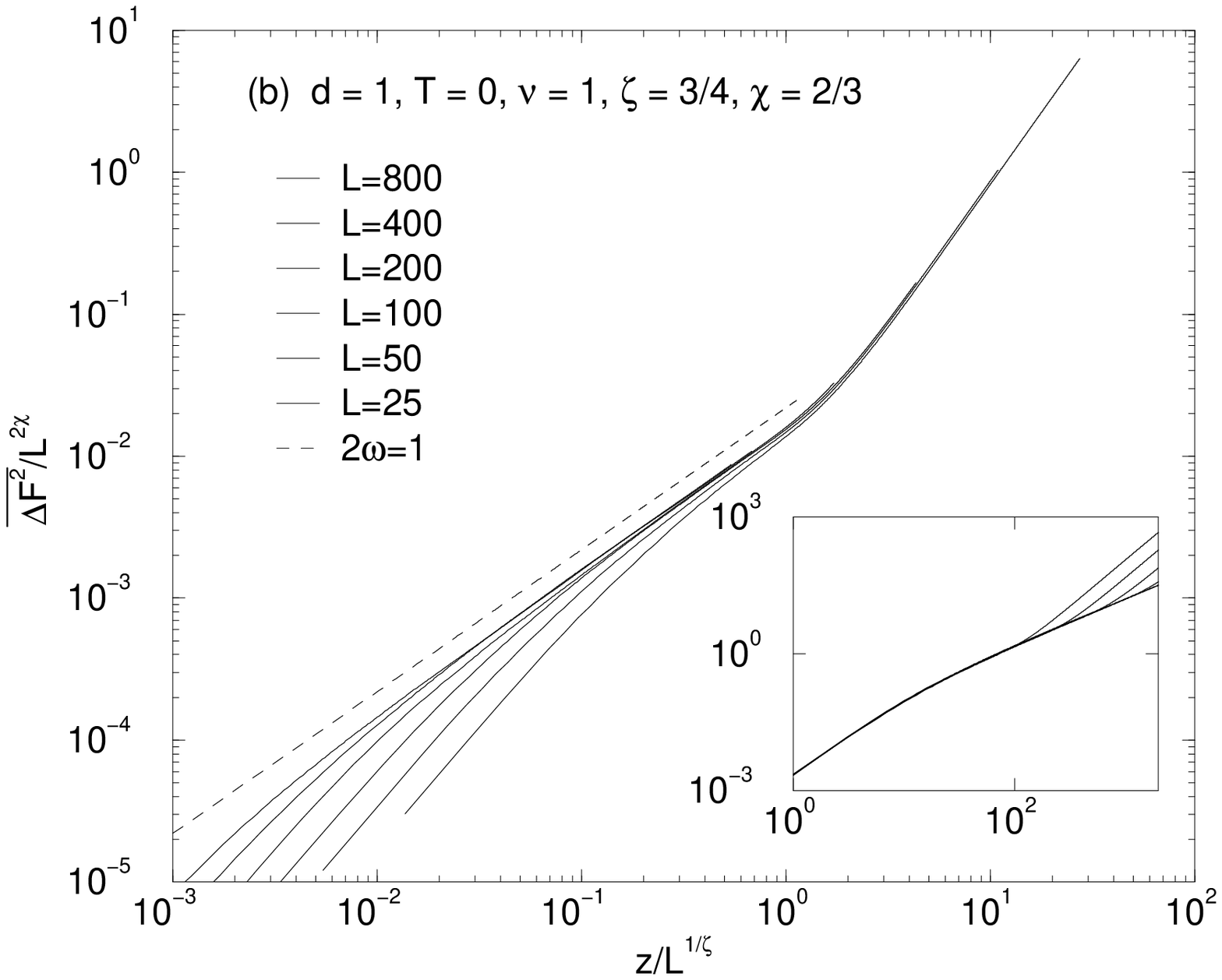} }
\caption{
(a) Scaling collapse according to \Eq{eq:collapse-x2} of mean square
fluctuations of the endpoint position of a flux line in (1+1)
dimensions for different system sizes at $T=0$.  The topmost curve
corresponds to the largest system, etc.  The statistical errors are
small and hardly visible in the figures here and below.  A clear power
law behavior with exponent $2\zeta = 3/2$ (dashed line) is
observed. The inset shows the unrescaled data, i.e., $\ave{\br^2}$ vs
$z$.
(b) Scaling collapse of the energy fluctuations of the optimal path.
The dashed line is a power law with exponent $2\omega=1$.  The inset
shows the unrescaled data, i.e., $\ave{\Delta F^2}$ vs $z$.
}
\label{fig:x2}
\label{fig:f}
\end{figure*}

\begin{figure*}
\centerline{
\epsfxsize= 0.5 \linewidth \epsfbox{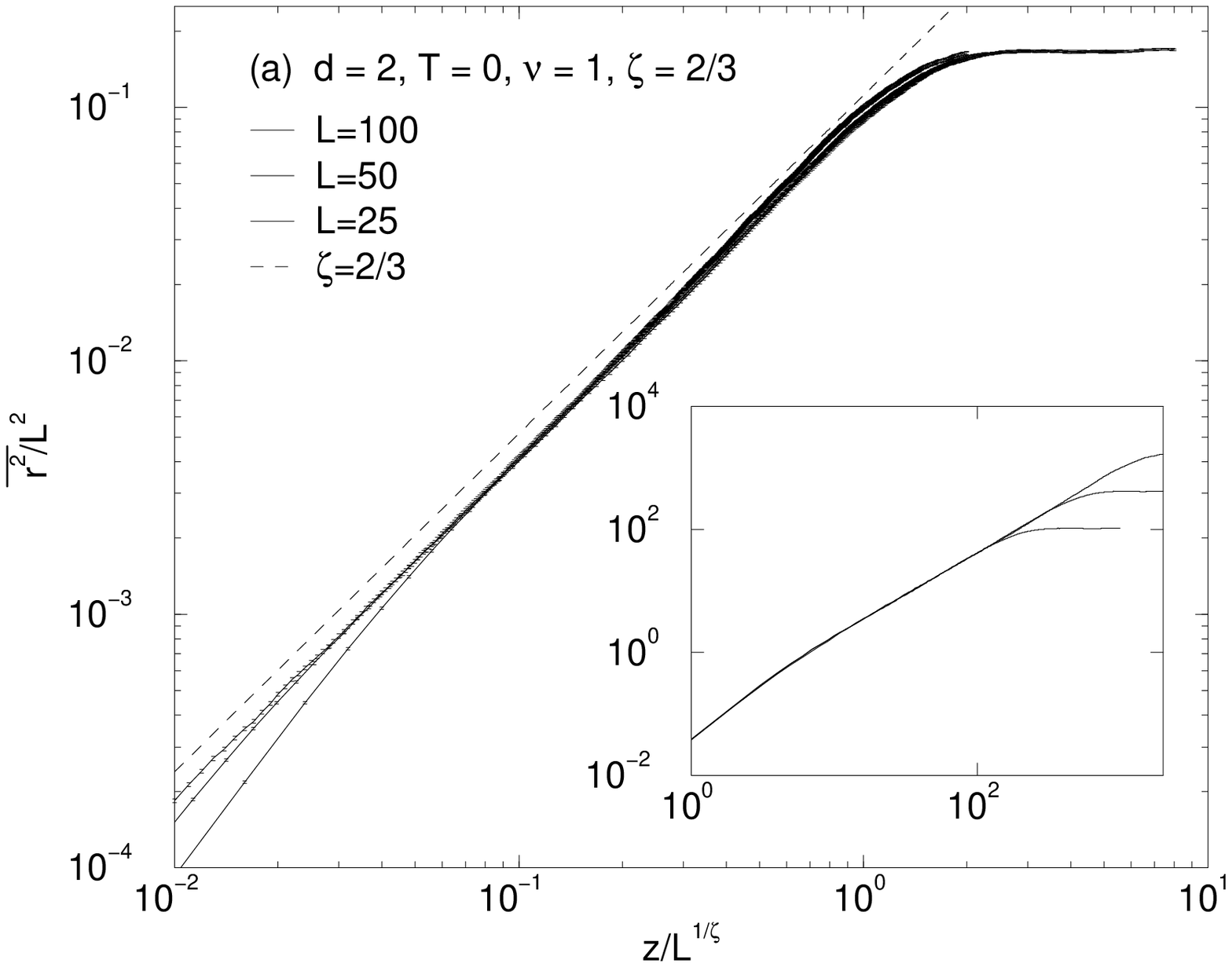}
\epsfxsize= 0.5 \linewidth \epsfbox{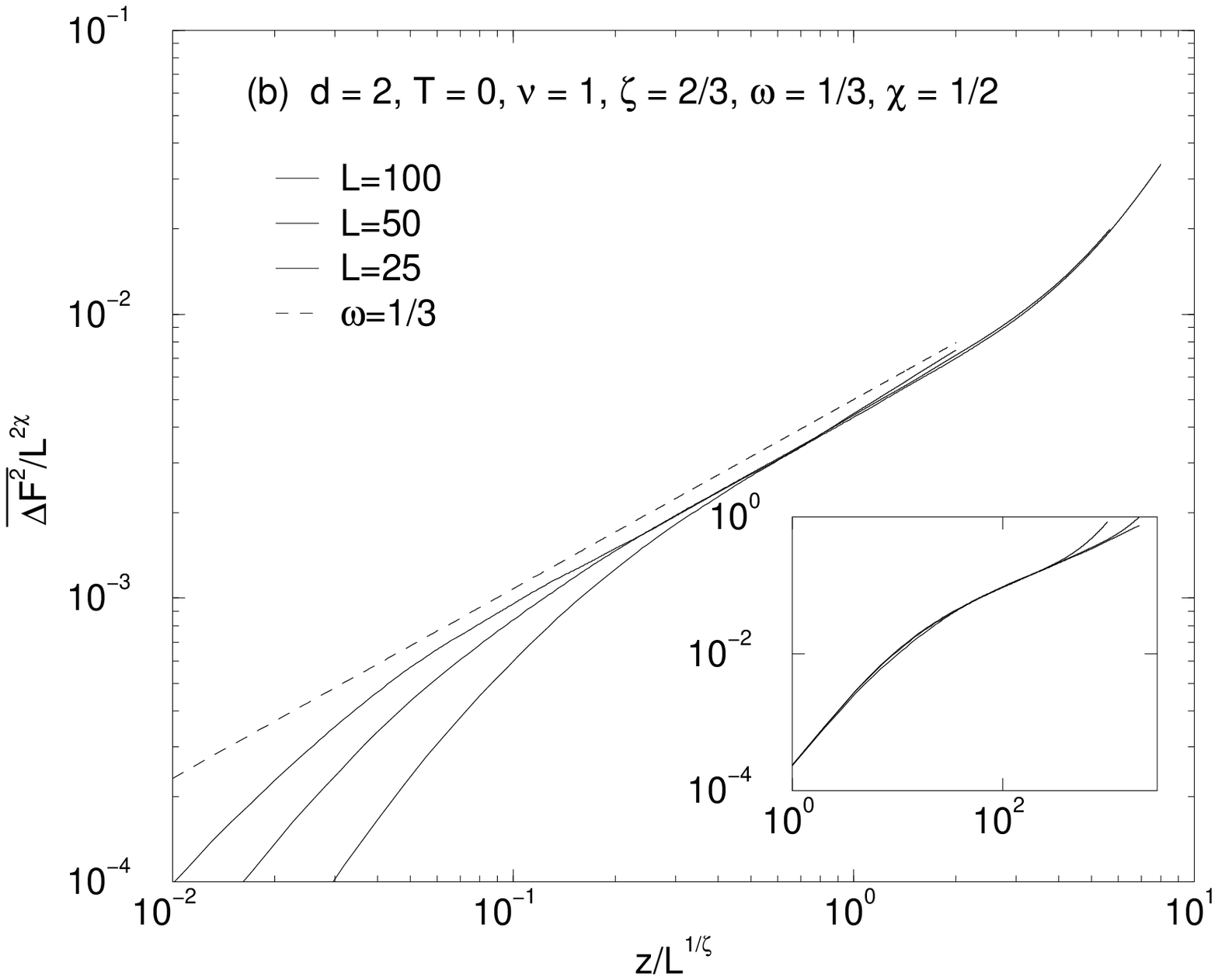} }
\caption{
(a) Scaling collapse of mean square fluctuations of the endpoint
position of a flux line in (2+1) dimensions for different system sizes
at $T=0$.  A power law behavior with exponent $2\zeta = 4/3$ (dashed
line) is observed. The inset shows the unrescaled data, i.e.,
$\ave{\br^2}$ vs $z$.
(b) Scaling collapse of the energy fluctuations of the optimal path.
The dashed line is a power law with exponent $2\omega=2/3$.  The inset
shows the unrescaled data, i.e., $\ave{\Delta F^2}$ vs $z$.
}
\label{fig:x2-3D}
\label{fig:f-3D}
\end{figure*}

The inset of \Fig{fig:x2}(a) shows the mean square positional
fluctuations as a function of $z$ for $d=1$ and $\nu=1$ at zero
temperature.  Note that the curves saturate for large $z$, consistent
with expectations from \Eq{eq:collapse-x2}.  The main part of the
figure shows a scaling collapse of the same data using the theoretical
value of $\zeta=3/4$.  In \Fig{fig:f}(b) we show a scaling collapse of
the energy fluctuations using $\omega=1/2$.  Here there are clearly
visible corrections to scaling for small $z$, and no power-law
behavior is observed for the smallest system sizes.  However, as $L$
is increased, the range over which scaling occurs increases.  In
\Fig{fig:x2-3D}(a) and \ref{fig:f-3D}(b) we show similar results for
$d=2$ and $\nu=1$, where \Eq{eq:exponents} predicts $\zeta=2/3$ and
$\omega=1/3$, and again the agreement is excellent.

\begin{figure*}
\centerline{ 
\epsfxsize= 0.5 \linewidth \epsfbox{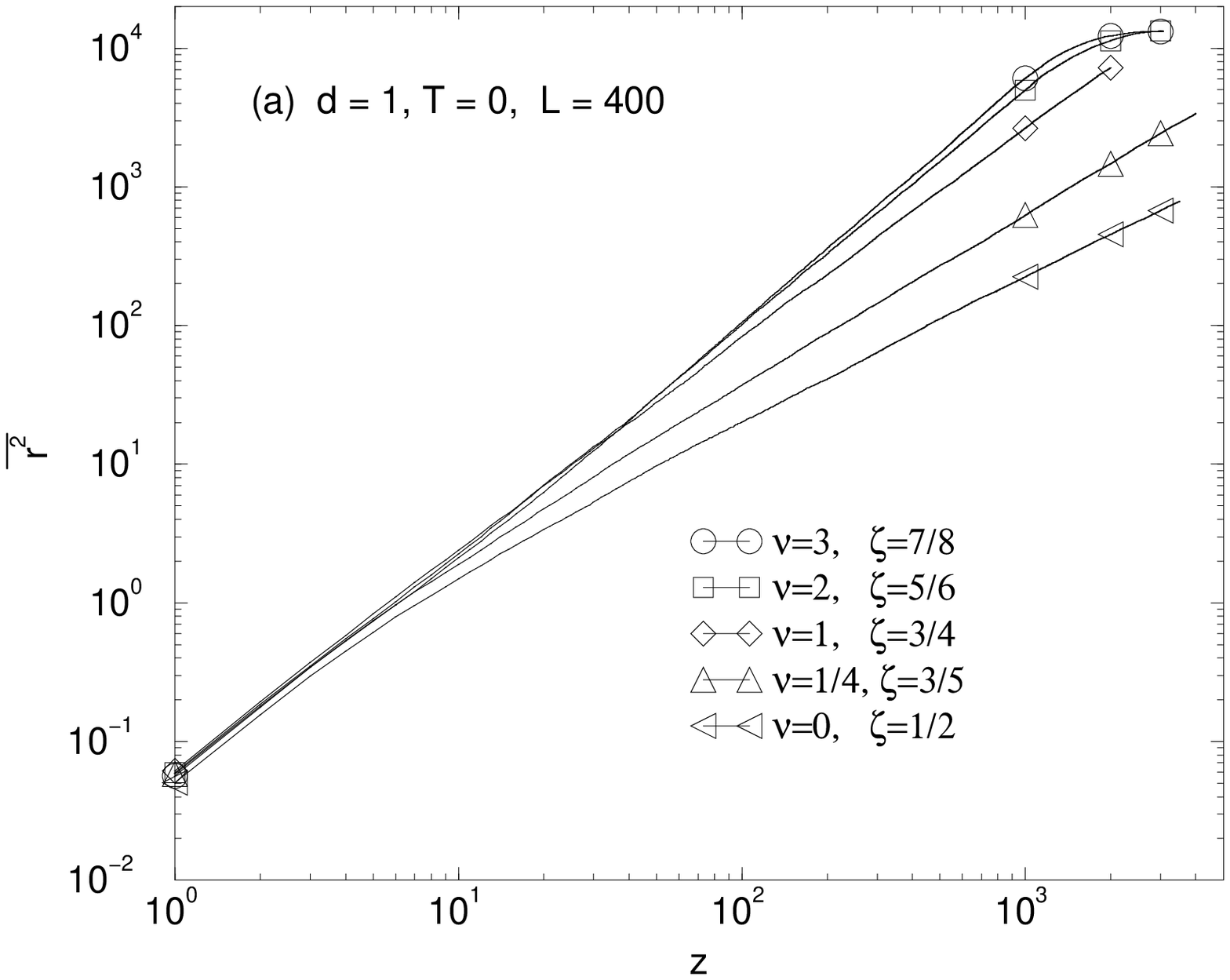}
\epsfxsize= 0.5 \linewidth \epsfbox{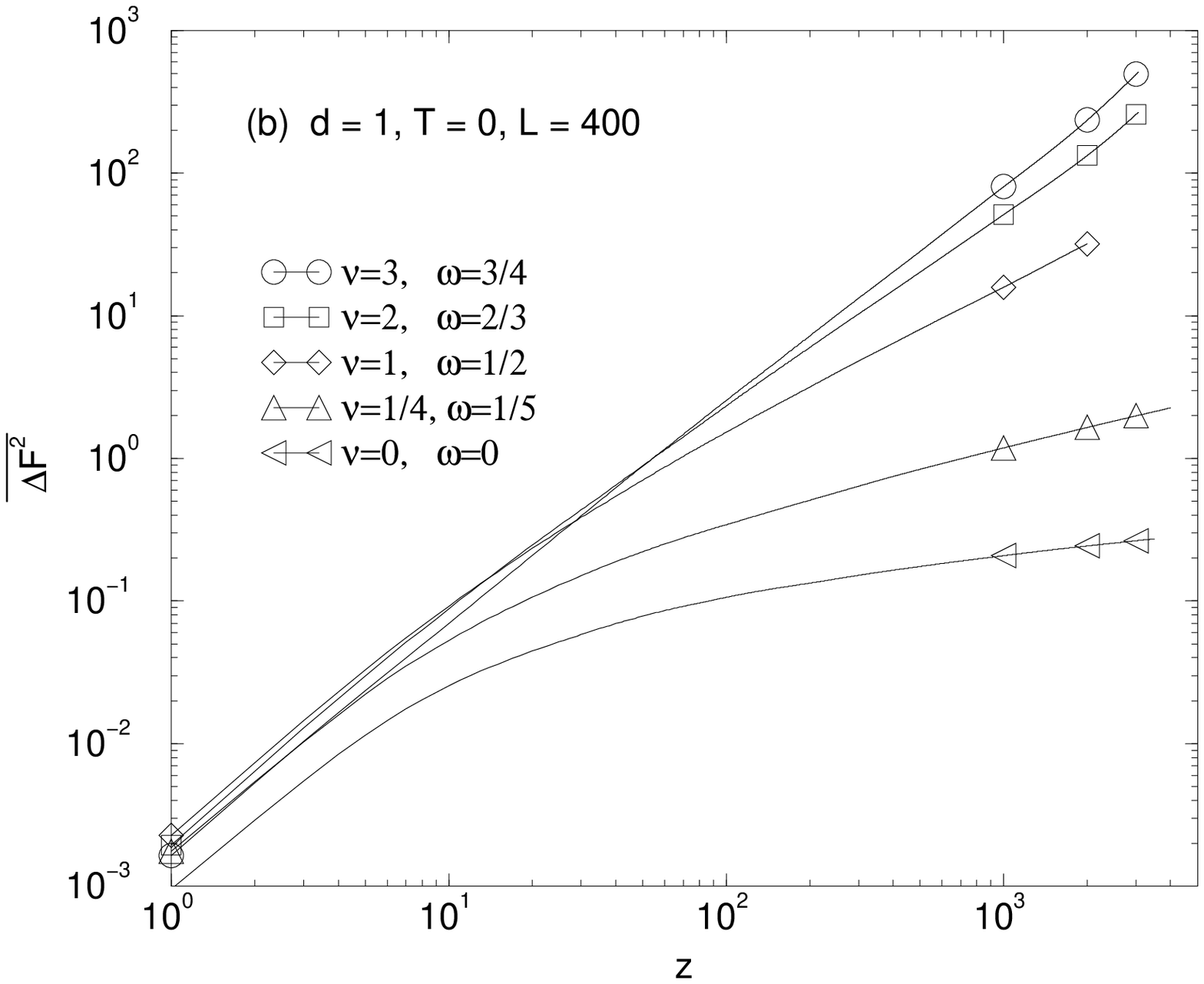} }
\caption{
(a) Mean square positional fluctuations for several different pinning
energy distributions, \Eq{eq:P(u)}, parameterized by $\nu$.  Power-law
fits (not shown) to the straight parts of the curves give exponents
that agree with \Eq{eq:exponents} to within $1\%$.
(b) Energy fluctuations for different $\nu$.
}
\label{fig:x2-nu}
\end{figure*}

To test the dependence of the exponents on the tail of the pinning
energy distribution we calculate the mean square positional
fluctuations [\Fig{fig:x2-nu}(a)] and the energy fluctuations
[\Fig{fig:x2-nu}(b)] for several different values of $\nu$.  For large
$z$ the data is accurately described by power laws with exponents
given by \Eq{eq:exponents}.  It is also possible to collapse the data
for different system sizes in each of these cases in a similar way to
\Fig{fig:x2}.

\begin{figure*}
\centerline{
\epsfxsize= 0.5\linewidth \epsfbox{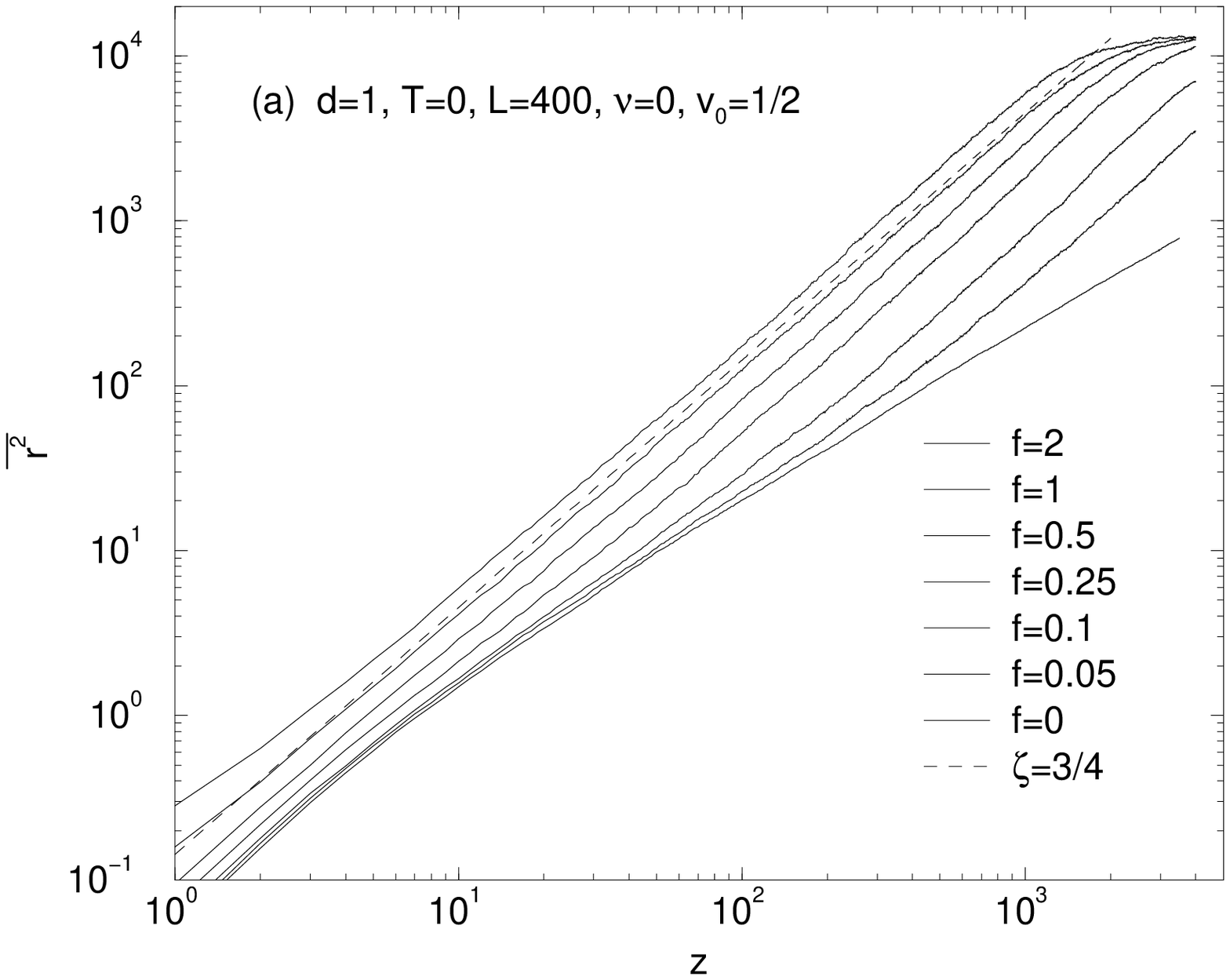} 
\epsfxsize= 0.5\linewidth \epsfbox{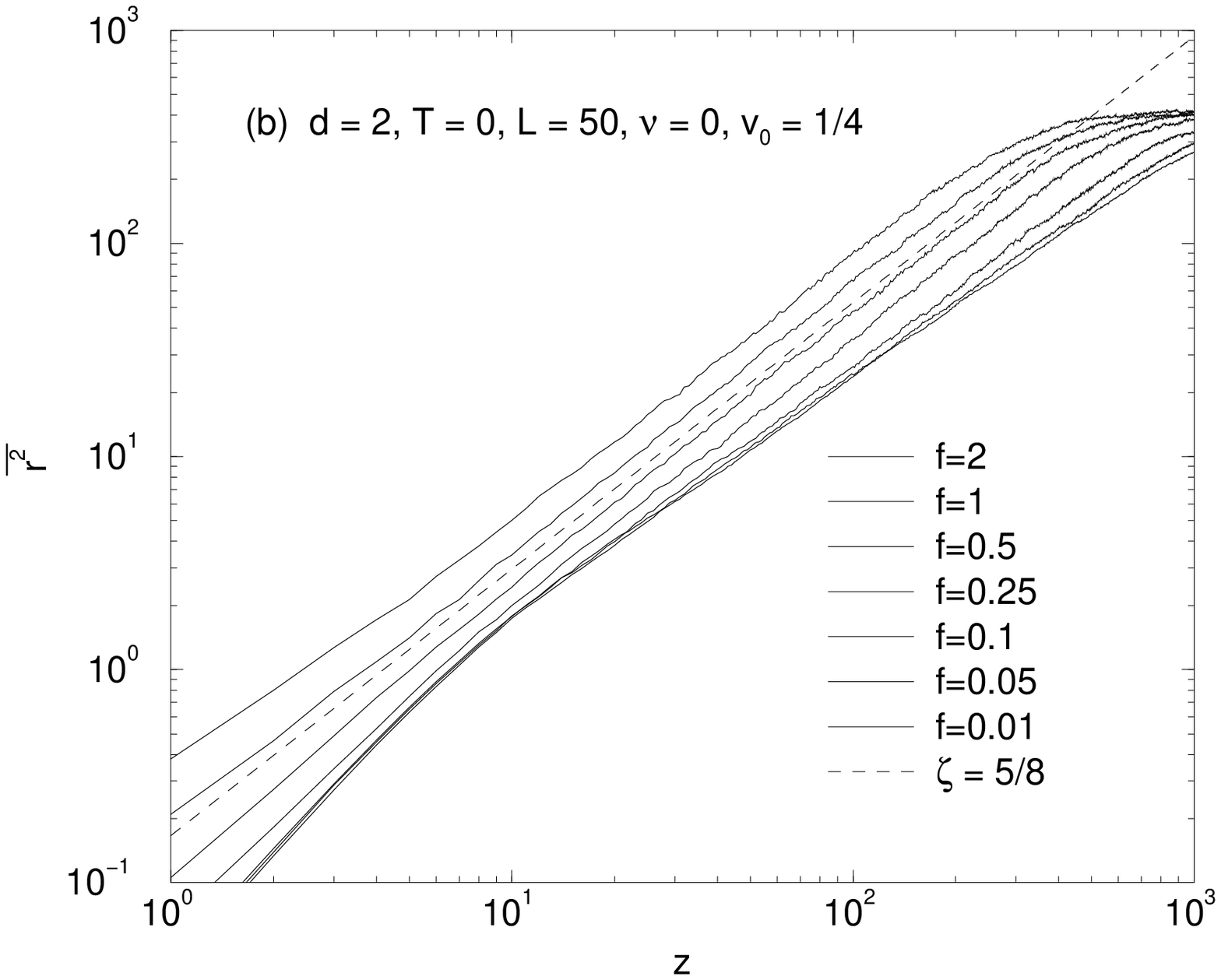} }
\caption{
Positional fluctuations for fragmented splayed columnar defects in (a)
(1+1) dimensions and (b) (2+1).  As the strenght of the fragmentation
increases the wandering exponent changes from $\zeta=1/2$ to a larger
value.  The dashed lines are power laws with exponents $\zeta=3/4$ and
$\zeta=5/8$, respectively.
}
\label{fig:x2-frag}
\end{figure*}

In \Fig{fig:x2-frag} we show results for fragmented splayed columnar
defects.  In (a) we plot the positional fluctuations for $d=1$,
$\nu=0$, for increasing strengths of the fragmentation $f$.  Since
$\nu<d$ fragmentation should be relevant, and already for small values
of $f$ we see deviations from the $f=0$ result, where $\zeta=1/2$.  As
$f$ is increased further $\zeta$ increases and stabilizes to a larger
value.  For large values of $f$ a power-law fit gives $\zeta\approx
3/4$ in agreement with \Eq{eq:frag}.  In (b) we plot the same quantity
for $d=2$.  For these parameters the scaling crosses over to that of
point disorder with $\zeta\approx 5/8$.
Thus, fragmentation seems to lead to point-like wandering for $d=2$,
while the results for $d=1$ are consistent with the second weak
fragmentation scenario of Sec.~\ref{sec:Fragmentation}.  For $\nu>d$
weak fragmentation should be irrelevant, and we do find that the
exponents for homogenous columns, \Eq{eq:exponents}, are much more
stable for small $f$ than they are for $\nu<d$.  For large $f$ the
behavior is still dominated by fragmentation on short length scales.
An accurate verification of the crossover scale, \Eq{eq:cross-over},
is very difficult with the limited system sizes we have studied, and
has not been attempted.

\begin{figure*}
\centerline{
\epsfxsize= 0.5\linewidth \epsfbox{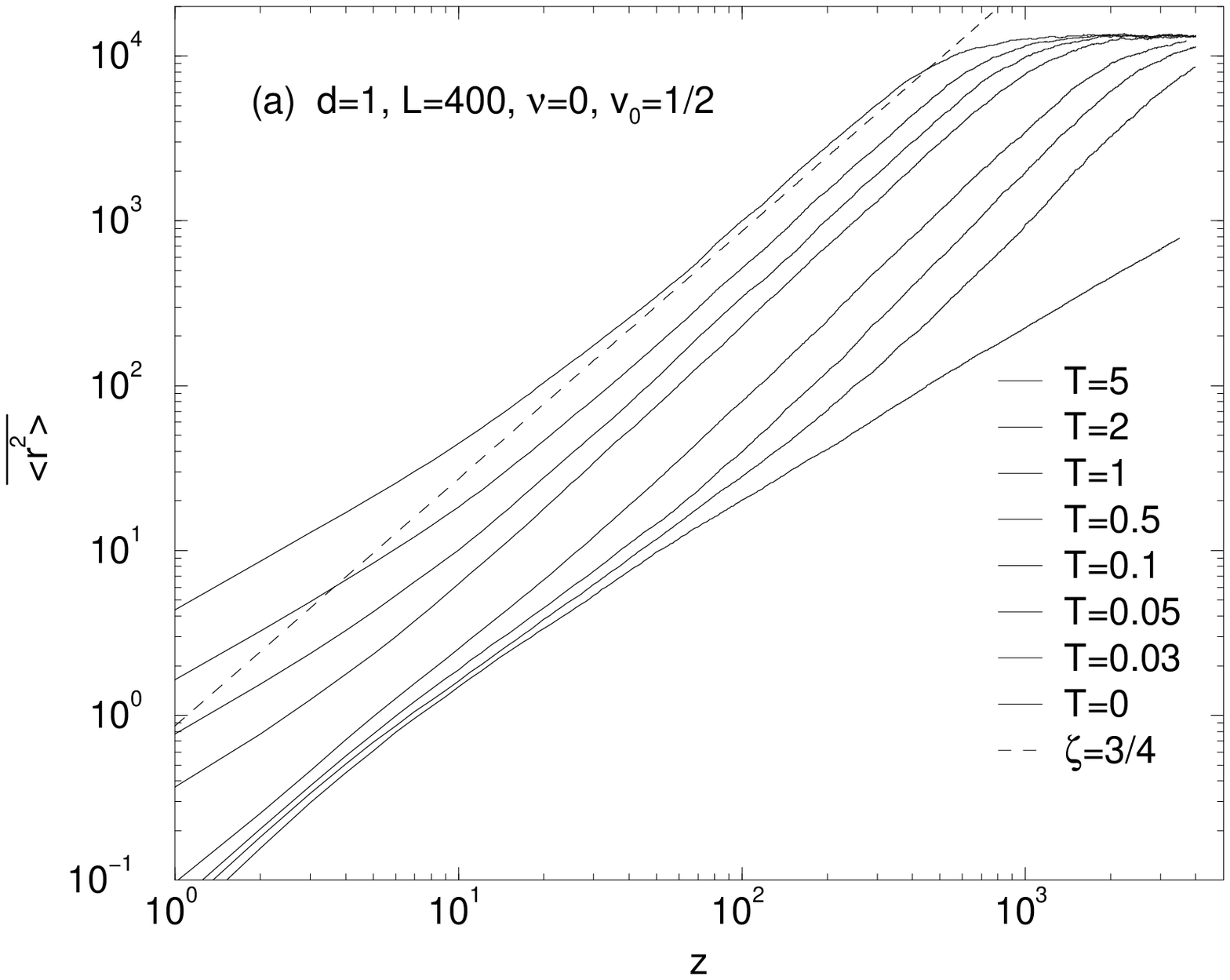}
\epsfxsize= 0.5\linewidth \epsfbox{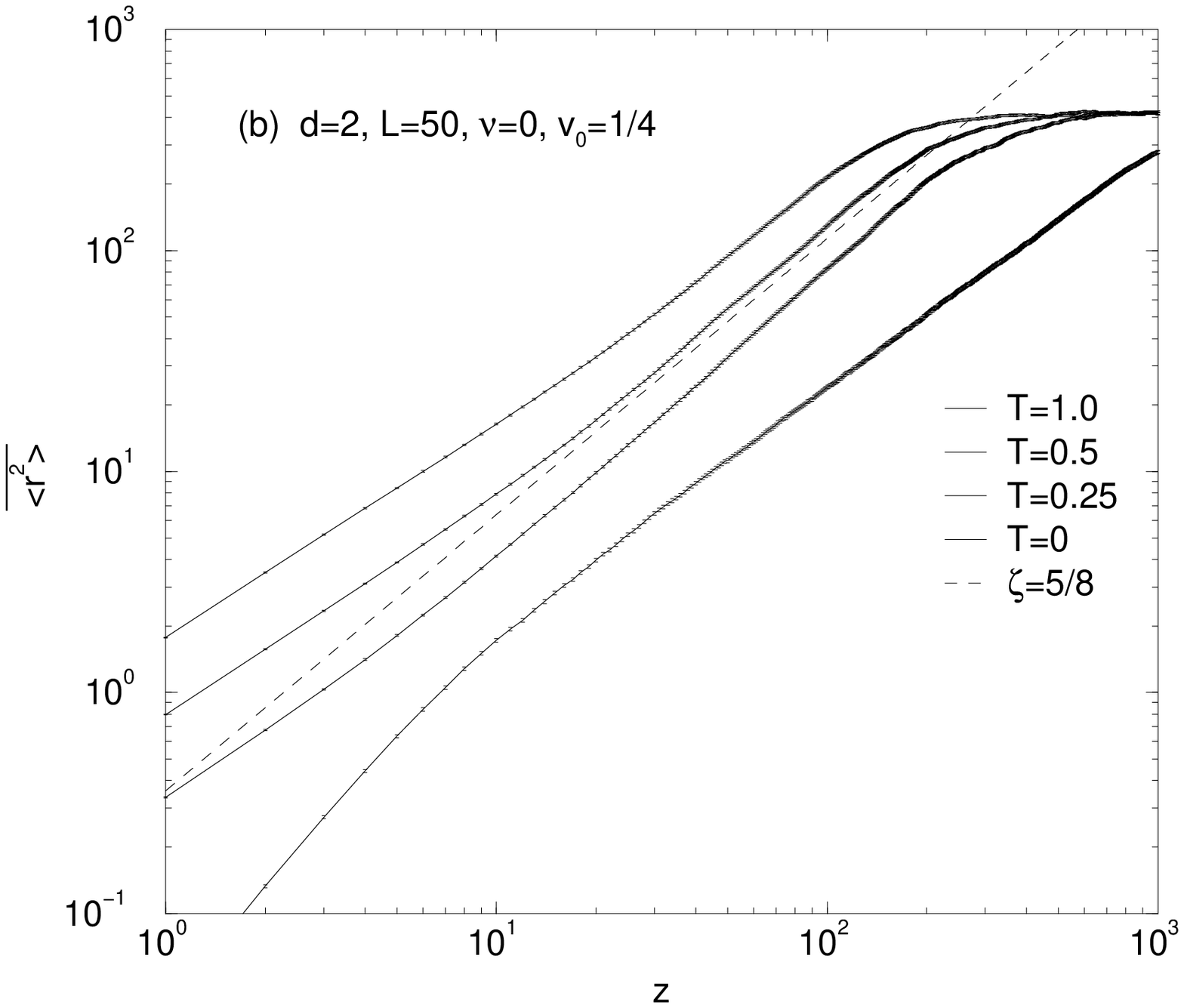} }
\caption{
Positional fluctuations for different temperatures in (a) (1+1)
dimensions and (b) (2+1).  The dashed lines are power laws with
exponents $\zeta=3/4$ and $\zeta=5/8$, respectively.
}
\label{fig:x2-T}
\end{figure*}

Finally, we explore the influence of finite temperature in
\Fig{fig:x2-T}.  As proposed in Sec.~\ref{sec:Finite-T} the scaling
shows strong resemblance to that of fragmented columnar defects,
compare \Fig{fig:x2-frag}.  In (1+1) dimensions [\Fig{fig:x2-T}(a)] a
powerlaw fit gives $\zeta\approx 0.75 - 0.8$, which is consistent with
$3/4$, the value for fragmented columnar defects.  In (2+1) dimensions
[(b)] we get instead $\zeta \approx 0.6 - 0.65$, consistent with the
value for point disorder, $\zeta \approx 5/8$.

\section{Future Directions}

\label{sec:future}

Our results indicate that the values of the wandering and energy
exponents depend sensitively on the low energy properties of the
pinning energy distribution.  In experiments one might therefore
expect these exponents to depend on the method of sample preparation.
A detailed comparison with experiments would require information about
the pinning energy distribution of the columnar defects for the
particular samples studied.  In case the columnar defects are
fragmented (and $\nu<d$) this dependence should disappear and
universality be restored.  The prospect of having a phase transition
between the two fragmentation scenarios of Sec~\ref{sec:Fragmentation}
is interesting and deserves further study.

In this paper we have focused our attention on the properties of
single flux lines in superconductors with splayed columnar defects.
At high vortex densities, i.e., at high magnetic fields, interactions
between vortices can no longer be neglected and the problem becomes
significantly more complicated.  
It would be interesting, although not easy, to extend both the
analytic arguments and the transfer matrix calculations to the case of
many interacting vortex lines.
We do, however, expect that the sensitivity to the low energy tails of
the pinning energy distribution should go away as the vortex density
is increased, since the low energy pins will all be occupied in this
case.

\section{Acknowledgment}

One of us (DRN) would like to acknowledge helpful conversations with
T.\ Hwa, and JL thanks E.\ Frey and Kihong Kim for stimulating
discussions.  This work was supported by The Swedish Foundation for
International Cooperation in Research and Higher Education (STINT),
the NSF through Grant No.\ DMR97-14725 and through the Harvard MRSEC
via Grant No.\ DMR98-09363.

\end{document}